\documentclass[conference,letterpaper]{IEEEtran}

\usepackage[utf8]{inputenc}
\usepackage[T1]{fontenc}
\usepackage{url}
\usepackage{ifthen}
\usepackage{cite}
\usepackage{amsfonts}       
\usepackage[cmex10]{amsmath} 
\usepackage{graphicx}
\usepackage{tikz}
\usepackage{subfig}
\usepackage{bm}
\usepackage{multirow}
\usepackage{algorithmic}

\begin{document}
\title{Protograph-Based Design for QC Polar Codes}

\author{%
  \IEEEauthorblockN{Toshiaki Koike-Akino and Ye Wang}
  \IEEEauthorblockA{Mitsubishi Electric Research Laboratories (MERL),
    201 Broadway, Cambridge, MA 02139, USA.\\
    Email: \{koike, yewang\}@merl.com}
}

\maketitle

\begin{abstract}

We propose a new family of polar coding which realizes high coding gain, low complexity, and high throughput by introducing a protograph-based design.
The proposed technique called as quasi-cyclic (QC) polar codes can be highly parallelized without sacrificing decoding complexity.
We analyze short cycles in the protograph polar codes and develop a design method to increase the girth.
Our approach can resolve the long-standing unsolved problem that belief propagation (BP) decoding does not work well for polar codes due to the inherently short cycles.
We demonstrate that a high lifting factor of QC polar codes can improve the performance and that QC polar codes with BP decoding can outperform conventional polar codes with state-of-the-art list decoding.
Moreover, we show that a greedy pruning method can improve the performance-complexity trade-off.

\end{abstract}

\section{Introduction}

Capacity-approaching forward error correction (FEC) based on low-density parity-check (LDPC) codes\cite{Smith-ite, Kudekar-CC, Koike-ite, Liva-PEXIT, Chang-PEXIT, Wei-NB, EXIT, SEXIT, Thorpe-proto03, Thorpe-proto04, Fossorier-girth, Tanner-girth, Kim-girth, Wang-girth} have made a great contribution to increasing data rates of wireless and optical communication systems.
However, the pursuit of high FEC performance has led to a significant increase in power consumption and circuit size.
Hence, attaining a good trade-off between performance and computational complexity is of great importance.
In addition, recent high-performance LDPC codes usually require very large codeword lengths,
whereas shorter FEC codes are preferred\cite{Liva-short} for latency-constrained systems, such as Internet-of-Things (IoT) applications.

Polar codes\cite{Arikan-polar, Tal-list, Shin-polar, Tal-design, Arikan-sys, Sarkis-sys, Li-list, Seidl-polar, Mori-DE, Trifonov-GA, He-beta, Elkelesh-GA, Ebada-DL, Alamdar-SSC, Doan, Elkelesh-BPL, Presman-mixed, Mori-NB, Gabry-kernel, relaxedPolar, Koike-JLT18, Koike-ICC18, Zhang-CRC, Wang-RS, Li-RM, Bourassa} have drawn much attention as alternative capacity-approaching codes in place of LDPC codes for short block lengths, in particular for the fifth-generation (5G) networks.
Besides encoder design methods\cite{Tal-design, Mori-DE, Trifonov-GA, He-beta, Elkelesh-GA, Ebada-DL},
a number of decoder algorithms were developed\cite{Alamdar-SSC, Doan, Elkelesh-BPL}.
With successive cancellation list (SCL) decoding\cite{Tal-list}, polar codes can be highly competitive with state-of-the-art LDPC codes.
To date, various extended versions based on polar coding have also been proposed in literature; e.g.,
nonbinary\cite{Mori-NB}, mixed-kernel\cite{Presman-mixed, Gabry-kernel}, irregular\cite{relaxedPolar, Koike-JLT18},
concatenated\cite{Zhang-CRC, Wang-RS, Li-RM}, convolutional\cite{Bourassa}, and turbo product coding\cite{Koike-ICC18}.

\begin{figure}[t]
 \centering
 \includegraphics[width=0.9\linewidth]{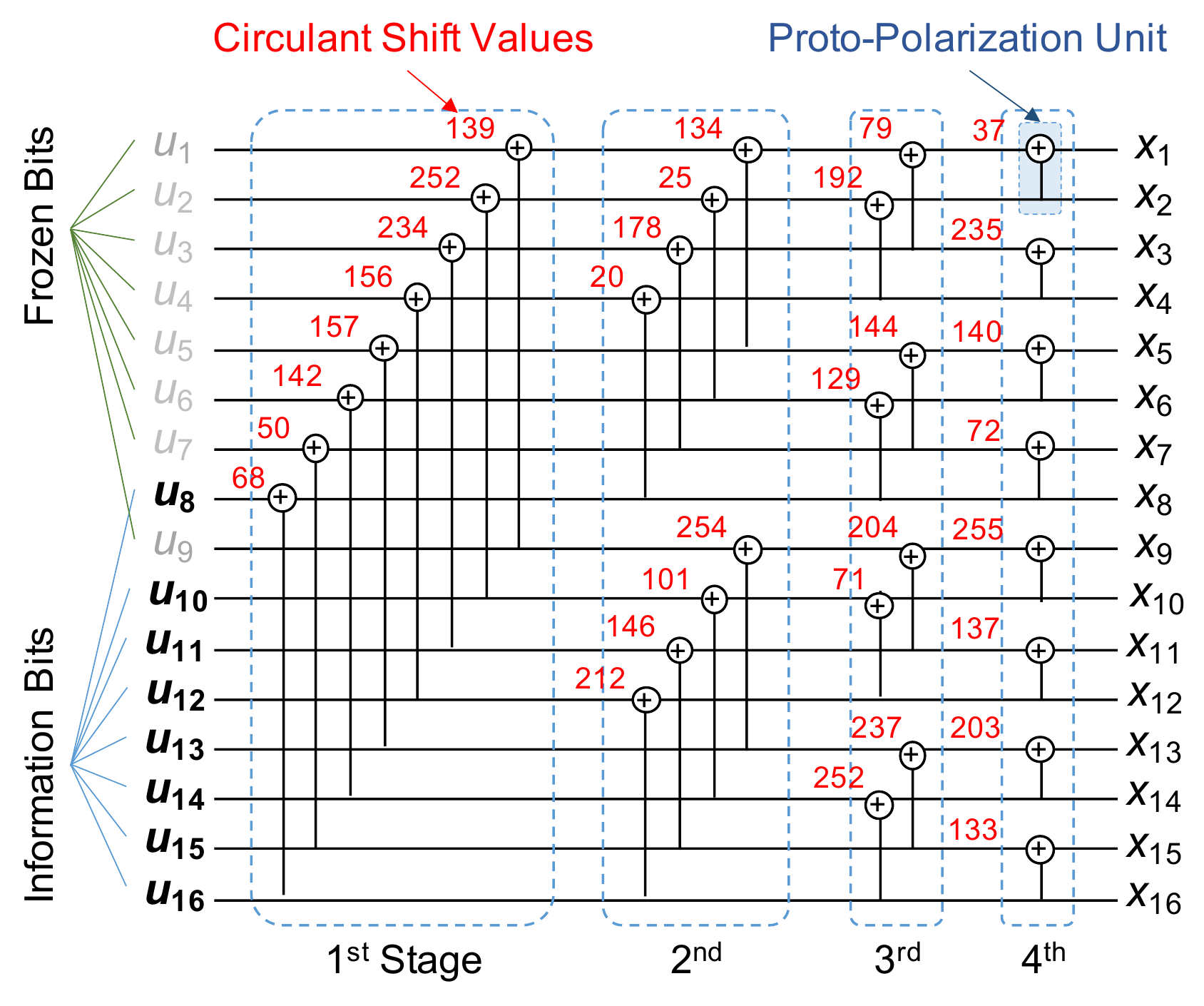}
 \caption{Four-stage polarization: QC polar codes $(2^4,2^3,2^8)$ having circulant shift values for $32$ proto-polarization units.}
 \label{fig:polar16}
\end{figure}

In this paper, we introduce a novel family of protograph-based polar codes, which we call quasi-cyclic (QC) polar codes having circulant permutation at proto-polarization units as illustrated in Fig.~\ref{fig:polar16}.
With a proper circulant shift value, we show that the QC polar codes can eliminate short cycles in the code graph, which achieves a remarkable breakthrough toward resolving the long-standing issue that the belief-propagation (BP) decoding does not perform well for polar codes.
In the QC polar codes, highly parallel short polar codes are coupled to achieve performance comparable to longer polar codes while maintaining the computational complexity as low as that of short polar codes.
The contributions of this paper are summarized as follows:
\begin{itemize}
 \item {\bf{Protograph-based polar codes}}:
       We propose a new family of protograph polar codes.
       To the best of the authors' knowledge, the concept of protograph codes has never been applied to
       such non-parity-check codes.
 \item {\bf{QC polar codes}}:
       We introduce highly parallelizable QC polar codes, a special case of protograph, with circulant permutations at the proto-polarization units.
 \item {\bf{Complexity analysis}}:
       We show that the computational complexity of the proposed QC polar codes can be significantly decreased with a protograph lifting factor.
 \item {\bf{Girth analysis}}:
       We analyze short cycles of the protograph polar codes, and develop a design method to increase the girth.
       Eliminating short cycles enables BP decoding to properly work for the QC polar codes.
 \item {\bf{State-of-the-art performance}}:
       We demonstrate that the QC polar codes with shallow polarization can achieve competitive performance of deep polarization codes.
 \item {\bf{Irregular pruning}}:
       Further complexity reduction and performance improvement are shown with irregular pruning of polarization to cut loops in the protograph.
\end{itemize}

\section{Basics of Polar Codes}

\subsection{Polar Encoding}

An $n$-stage polar code with $K$ information bits and $N=2^n$ encoded bits uses an $N\times N$ generator matrix $\boldsymbol{G}^{ \otimes n}$ for encoding, where
$[\cdot]^{\otimes n}$ denotes the $n$-fold Kronecker power and $\boldsymbol{G}$ is a binary kernel matrix defined as
\begin{align}
 \boldsymbol{G} &=
  \begin{bmatrix}
   1 & 1 \\
   0 & 1 \\
  \end{bmatrix}
 .
\end{align}
Let ${\boldsymbol{u}} = [{u_1},{u_2},\ldots,{u_N}]^\mathrm{T}$ and
${\boldsymbol{x}}= [{x_1},{x_2},\ldots,{x_N}]^\mathrm{T}$ respectively denote the column vectors of input bits and encoded bits.
The codeword (for non-systematic codes) is given
by $\boldsymbol{x} = \boldsymbol{G}^{\otimes n} \boldsymbol{B} \boldsymbol{u}$, where
the matrix multiplications are carried out over the binary field (i.e., modulo-$2$ arithmetic),
and $\boldsymbol{B}$ denotes an $N \times N$ bit-reversal permutation matrix\cite{Arikan-polar}.
Due to the nature of the Kronecker product, polar encoding and decoding
can be performed at a complexity on the order of $\mathcal{O}(N \log_2 N)$. 
The multi-stage operation of the Kronecker products gives rise to the so-called
polarization phenomenon to approach capacity in arbitrary channels\cite{Arikan-polar}.

The polar coding maps the information bits to the $K$ most reliable locations in $\boldsymbol{u}$.
The remaining $N-K$ input bits are frozen bits, known to both encoder and decoder.
We use $\mathbb{K}$ and $\bar{\mathbb{K}}$ to denote the subsets of $\{1, 2, \ldots, N\}$ that correspond to the information bit and frozen bit locations, respectively.
The lowest reliability can be selected to be in $\bar{\mathbb{K}}$ for frozen bits, e.g., by Bhattacharyya parameter\cite{Arikan-polar}, density evolution\cite{Mori-DE, Tal-design},
Gaussian approximation\cite{Trifonov-GA}, beta expansion\cite{He-beta}, genetic algorithm\cite{Elkelesh-GA}, and deep learning\cite{Ebada-DL}.

\subsection{Polar Decoding}

The original SC decoder\cite{Arikan-polar} proceeds sequentially over the bits, from $u_1$ to $u_N$.
For each index $i \in \{1,2, \ldots, N\}$, an estimate $\hat{u}_i$ for bit $u_i$ is made as follows.
If $i \notin \mathbb{K}$, then $\hat{u}_i$ is set to the known value of $u_i$, otherwise, when $i \in \mathbb{K}$, $\hat{u}_i$ is set to the most likely value for $u_i$ given the channel outputs,
assuming that the previous estimates $[\hat{u}_1, \hat{u}_2, \ldots, \hat{u}_{i-1}]$ are correct.
The SC decoding was improved by the SCL decoder\cite{Tal-list}, 
which proceeds similarly to the SC decoder, except that for each data bit index $i \in \mathbb{K}$, the decoder retains both possible estimates, $\hat{u}_i = 0$ and $\hat{u}_i = 1$, in subsequent decoding paths.
The list-decoding approach limits the number of paths to a fixed-size list $L$ of the most likely partial paths.
The combination of SCL decoding with an embedded cyclic redundancy check (CRC) to reject invalid paths yields significantly improved performance\cite{Tal-list, Zhang-CRC}.
Various other decoding algorithms were proposed in the literature, e.g., simplified SC decoding\cite{Alamdar-SSC}, neural SC decoding\cite{Doan}, and BP list decoding\cite{Elkelesh-BPL}.

\subsection{Computational Complexity}
\label{sec:comp}

It is known that short LDPC codes do not perform well as shown in~\cite{Liva-short}, where
 nonbinary (NB) coding can improve LDPC codes in the short-length regimes.
However, the computational complexity of NB-LDPC decoding is generally higher than binary counterparts, in particular for large Galois field sizes.
It is thus of great importance to realize low computational complexity in addition to high coding gain.
We evaluate the computational complexity of polar decoding and show that it is competitive with LDPC decoding.

\begin{figure}[t]
 \centering
 \includegraphics[width=\linewidth]{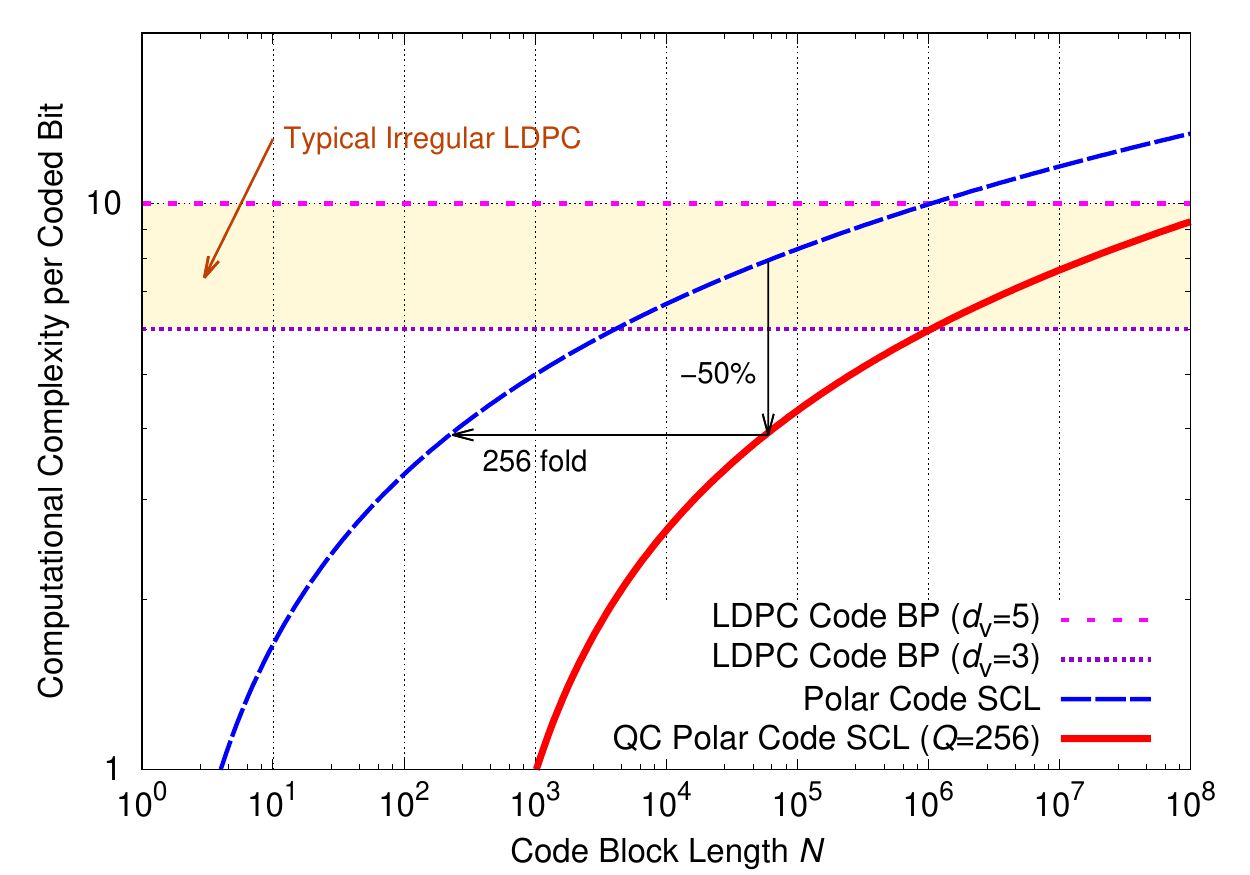}
 \caption{Computational complexity per coded bit as a function of block length $N$ for standard polar SCL decoding (per list), QC polar BP decoding and LDPC BP decoding (per iteration).}
 \label{fig:comp}
\end{figure}

The polar SCL decoding has a log-linear complexity; specifically, $\mathcal{O}[L N \log_2(N)/2]$) for a list size of $L$.
This nonlinearity is a major drawback in comparison to the linear complexity of LDPC BP decoding, i.e.,
$\mathcal{O}[2I d_\mathrm{v} N]$ where $d_\mathrm{v}$ denotes the average degree of variable nodes (VNs).
Note that the factor of $2I d_\mathrm{v}$ comes from the bidirectional message passing, whereas SCL decoding uses unidirectional message passing over $N\log_2(N)/2$ VNs.
Due to the nonlinear complexity, polar codes can eventually be less effective than LDPC codes as we increase the block lengths $N$.
However, it turns out to be an advantage when we aim to reduce the block sizes in order to decrease decoding latency.
This is illustrated in Fig.~\ref{fig:comp}, where complexity per coded bit (i.e., divided by $N$) is plotted as a function of block length $N$ for polar and LDPC decoding.
Because per-bit complexity is constant depending on average degree $d_\mathrm{v}$ for LDPC codes, there is no
motivation to decrease the block length.
In contrast, polar decoding becomes simpler when we reduce block sizes.
Remarkably, polar decoding will be more efficient than typical LDPC decoding at short block lengths of $N < 10^4$.
This promotes polar codes as a strong candidate for latency-critical systems.

Although the actual computational complexity may vary depending on hardware implementation,
most prototyping studies\cite{Sarkis-sys} have revealed that polar codes can compete favorably with LDPC codes in terms of complexity.
Note that the LDPC decoding is more complicated for higher rates because
the average check-node (CN) degree is larger, whereas polar codes have at most three degrees at CNs.
Nevertheless, polar SCL decoding is not amenable to parallel implementation.
In this paper, we propose a highly parallelizable polar code family whose complexity is $\mathcal{O}[LN\log_2(N/Q)/2]$ for a parallelism factor of $Q$.
From Fig.~\ref{fig:comp}, we can observe the significant advantage in its decoding complexity.
The details of our proposed QC polar codes will be described in the next section.

\section{Protograpah-Based QC Polar Codes}

\subsection{Protograph Codes}
\begin{figure}[t]
 \centering
 \includegraphics[width=\linewidth]{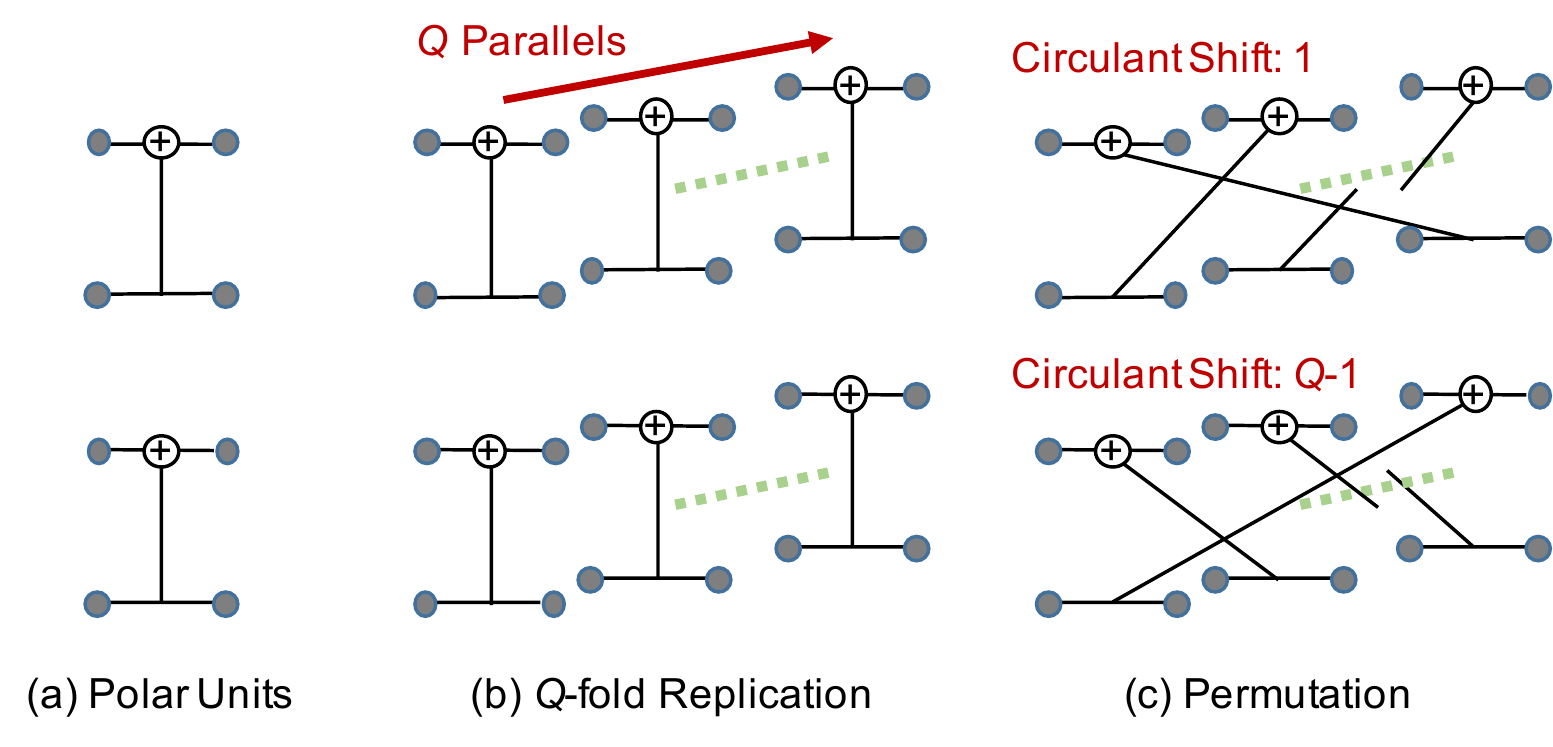}
 \caption{Lifting operation for proto-polarization units:
 (a) regular polar units,
 (b) replication of $Q$-parallel encoders,
 (c) permutation for interleaving intermediate encoding bits.}
 \label{fig:lifting}
\end{figure}

Thorpe\cite{Thorpe-proto03} introduced the concept of protograph codes, a class of LDPC codes constructed from a protograph in such a way that the $1$'s in the parity-check matrix are replaced by $(Q\times Q)$-permutation matrices and the $0$'s by $(Q\times Q)$-zero matrices.
The permutation size $Q$ is also called a lifting size.
If the permutation matrices are circulant, the protograph code reduces to a well-known QC LDPC code\cite{Thorpe-proto04}.
To the best of authors' knowledge, no studies have been reported for designing the protographs for polar codes.

Analogously in lifting operations of the parity-check matrix for LDPC codes, we replace the generator matrix of polar codes.
For example, the following generator matrix for $2$-stage polar codes is replaced with permutation matrices $\boldsymbol{P}_{i,j}$:
\begin{align}
 \boldsymbol{G}^{\otimes 2} &=
  \begin{bmatrix}
   1 & 1 & 1 & 1\\
   0 & 1 & 0 & 1\\
   0 & 0 & 1 & 1\\
   0 & 0 & 0 & 1\\
  \end{bmatrix}
 \mathop{\Longrightarrow}_{\mathrm{Lifting}}
  \begin{bmatrix}
   \boldsymbol{P}_{1,1} & \boldsymbol{P}_{1,2} & \boldsymbol{P}_{1,3} & \boldsymbol{P}_{1,4} \\
   \boldsymbol{0} & \boldsymbol{P}_{2,2} & \boldsymbol{0} & \boldsymbol{P}_{2,4}\\
   \boldsymbol{0} & \boldsymbol{0} & \boldsymbol{P}_{3,3} & \boldsymbol{P}_{3,4}\\
   \boldsymbol{0} & \boldsymbol{0} & \boldsymbol{0} & \boldsymbol{P}_{4,4}\\
  \end{bmatrix}
 ,\notag
\end{align}
where $\boldsymbol{0}$ is an all-zero matrix of size $Q\times Q$.
The simplest choice of permutation matrices is a weight-$1$ circulant matrix:
$
 \boldsymbol{P}_{i,j} =
 \boldsymbol{I}(s'_{i,j})
$,
where $\boldsymbol{I}(s)$ denotes the $s$th circulant permutation matrix obtained by cyclically right-shifting a $Q\times Q$ identity matrix by $s$ positions, and $s'_{i,j}$ is a shift value to design.
For this special case, we may call the protograph polar codes as QC polar codes.
It can be regarded as a generalized low-density generator matrix (LDGM) based on polar codes.

We consider a hardware-friendly lifting operation at each polarization stage with identity diagonal matrices $\boldsymbol{P}_{i,i} = \boldsymbol{I}(0)$.
Our lifting operation is illustrated in Fig.~\ref{fig:lifting}, where we replicate $Q$-parallel polar encoders and permute exclusive-or (XOR) incident bits among the parallel encoders.
Fig.~\ref{fig:polar16} shows an example of QC polar codes having a shift base matrix of size $n\times 2^{n-1}$ as follows:
\begin{align}
 \boldsymbol{S} &=
 \begin{bmatrix}
  139 & 252 & 234 & 156 & 157 & 142 & 50 & 68 \\
  134 & 25 & 178 & 20 & 254 & 101 & 146 & 212 \\
  79 & 192 & 144 & 129 & 204 & 71 & 237 & 252 \\
  37 & 235 & 140 & 72 & 255 & 137 & 203 & 133 \\
 \end{bmatrix}
 ,
 \label{eq:shift16}
\end{align}
whose $(i,j)$th shift value is assigned for the $j$th proto-polarization unit at the $i$th stage.
Note that the QC polar codes still hold most benefits of original polar codes such as structured encoding and decoding.
As we will discuss below, the QC polar codes have a number of remarkable features.

\subsection{High-Girth Design}

\begin{figure}[t]
 \centering
 \includegraphics[width=\linewidth]{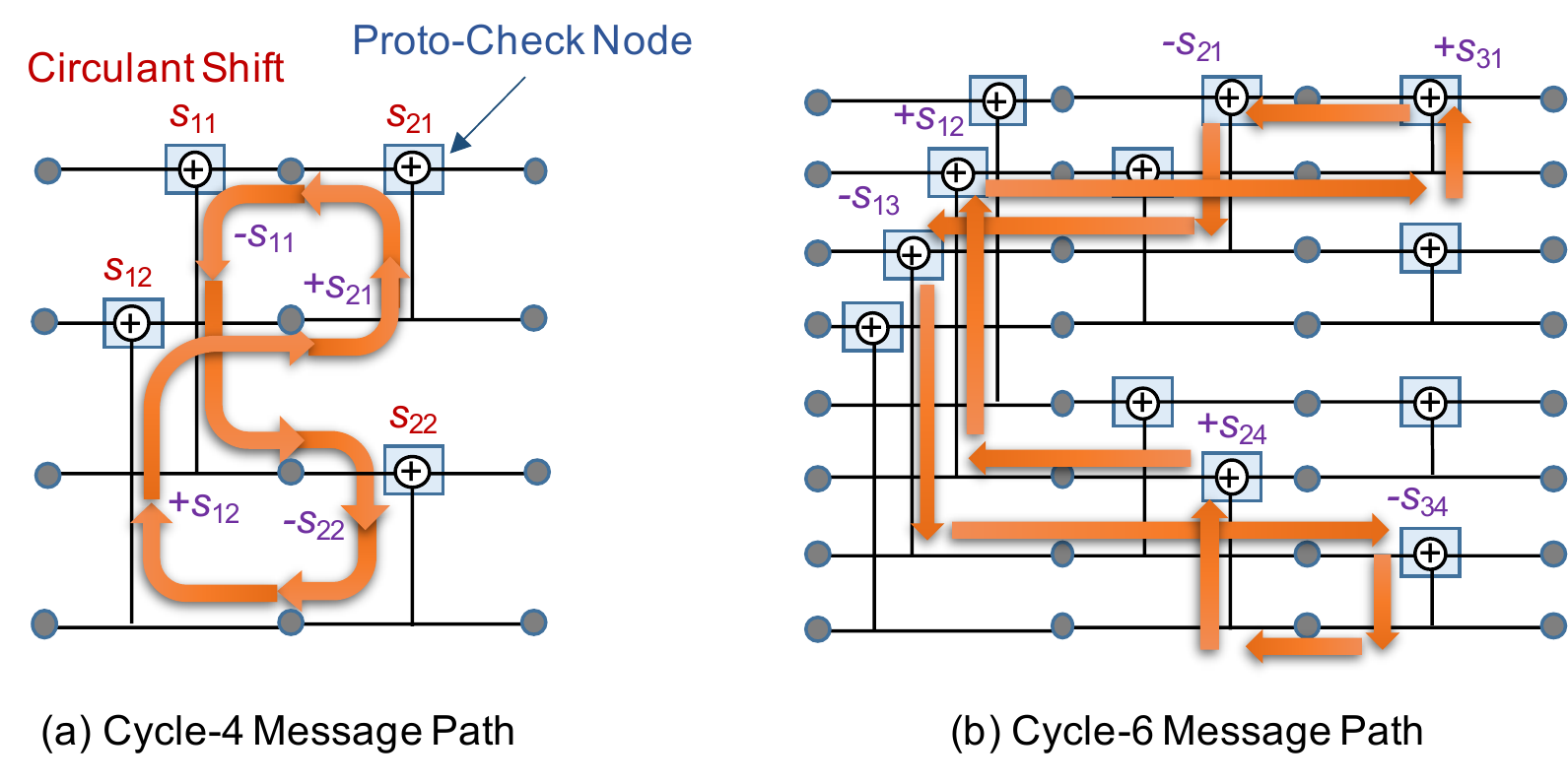}
 \caption{Short cycle examples in QC polar codes:
 (a) cycle-$4$ message passing loop,
 (b) cycle-$6$ message passing loop.}
 \label{fig:cycle}
\end{figure}

In order to achieve good performance, we shall design the shift values of QC polar codes.
One obviously poor choice is the case when we use all zeros for shifting, leading to mutually independent $Q$-parallel short polar codes without any coupling gain.
The protograph codes are often designed to achieve a high girth --- the ``girth'' of a code is the length of the shortest cycle in the code graph.
It is known\cite{Fossorier-girth} that the girth of any conventional QC LDPC code is upper bounded by $12$.
Tanner\cite{Tanner-girth} proposed a systematic way to optimize shift values to achieve girth-12.
It was further shown in~\cite{Kim-girth} that an irregular QC LDPC code can achieve a girth larger than 12.

For QC polar codes $(2^n, K, Q)$ of code length $N=2^nQ$, there are $n 2^{n-1}$ shift values to design as in (\ref{eq:shift16}).
Unfortunately, the factor graph of polar codes are inherently loopy and there exist a large number of short cycles as illustrated in Fig.~\ref{fig:cycle}.
Nonetheless, by optimizing shift values, we can increase the girth for QC polar codes with $Q>1$.
For example, the cycle-$4$ loop in Fig.~\ref{fig:cycle}(a) can be eliminated if the shift values satisfy the condition\cite{Wang-girth}:
\begin{align}
 {}-s_{1,1} - s_{2,2} + s_{1,2} + s_{2,1} \neq 0 \pmod Q,
 \label{eq:girth4}
\end{align}
where we accumulate shift values of all proto-CNs along the loop.
Note that the shift values are negated if the path goes downward.
This explains the long-lasting problem that the BP decoding performs very poorly for the conventional polar codes ($Q=1$), i.e., the accumulated shifts will be always zero, resulting in a small girth of $4$.
Our QC polar codes resolve this issue by maximizing the girth in the protograph.
Similarly, the cycle-$6$ loop in Fig.~\ref{fig:cycle}(b) can be removed if we can satisfy
\begin{align}
 {}-s_{1,3} -s_{3,4} + s_{2,4} + s_{1,2} + s_{3,1} - s_{2,1} \neq 0 \pmod Q.
 \notag
\end{align}
Note that irregular polar coding\cite{Koike-JLT18} is an alternative that could remove some, but not all, short-cycle loops.
We extended the greedy design method used for irregular polar coding to jointly optimize frozen bit locations and circulant shift values by means of protograph extrinsic information transfer (P-EXIT)\cite{Liva-PEXIT, Chang-PEXIT} and a hill-climbing girth maximization\cite{Wang-girth}.

\subsection{Error-Rate Performance}

\begin{figure}[t]
 \centering
 \includegraphics[width=\linewidth]{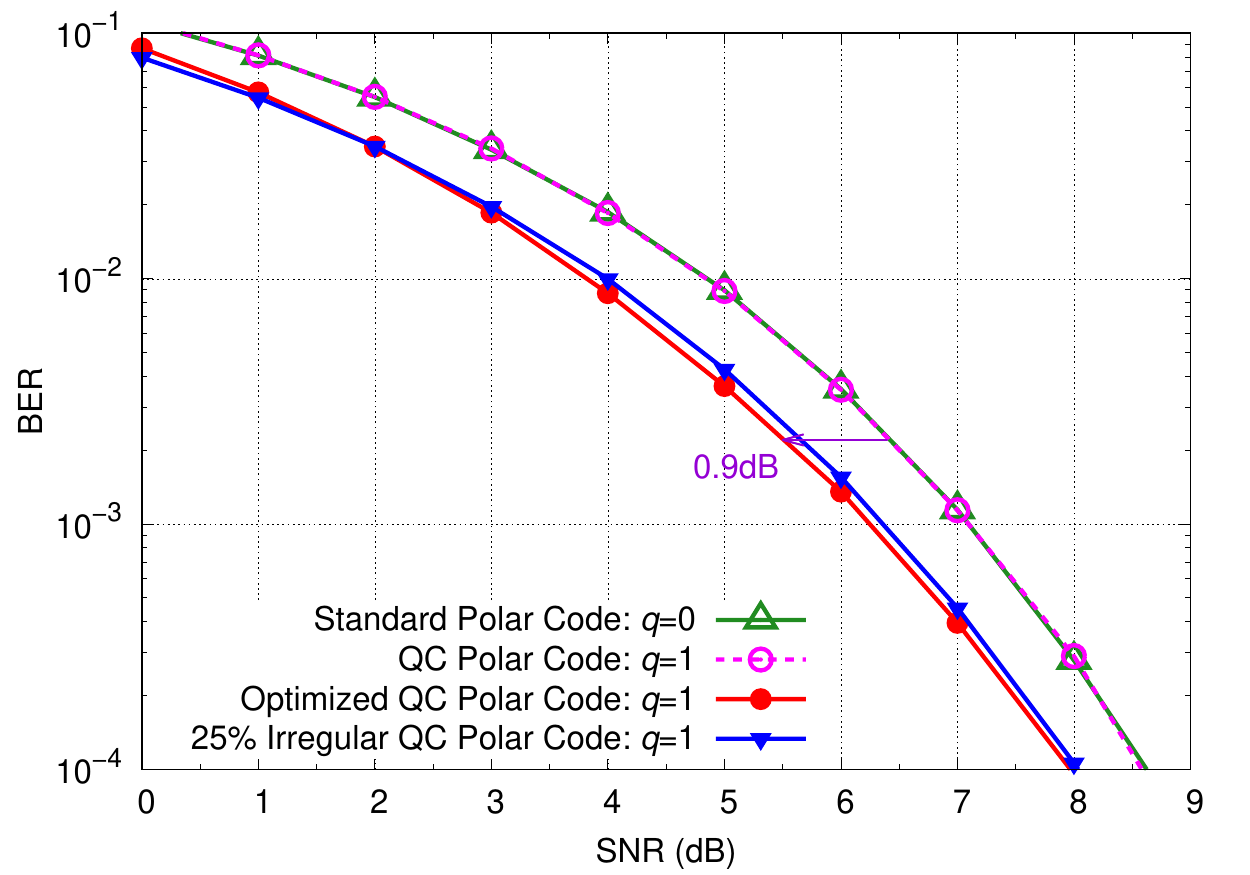}
 \caption{BER performance of $32$-iteration BP decoding for $2$-stage half-rate QC polar codes $(2^2,2^1,2^q)$.
 Frozen bit indications are $[1,1,0,0]$.}
 \label{fig:m2}
\end{figure}

Fig.~\ref{fig:m2} shows bit-error rate (BER) performance as a function of signal-to-noise ratio (SNR) for short polar codes $(2^n,2^{n-1},2^q)$ with $n=2$ polarization stages in additive white Gaussian noise channels.
We here use $32$-iteration BP decoding with two-way round-robin scheduling from the first to the last stages and its reversed direction alternatingly (parallel flooding updates per stage).
The first two bits $[u_1, u_2]$ are frozen.
The following shift base matrices are considered:
\begin{align}
 \begin{bmatrix}
  0 & 0 \\
  0 & 0 \\
 \end{bmatrix}
 ,\quad
 \begin{bmatrix}
  0 & 1 \\
  1 & 0 \\
 \end{bmatrix}
 ,\quad
 \begin{bmatrix}
  0 & 0 \\
  0 & 1 \\
 \end{bmatrix}
 ,\quad
 \begin{bmatrix}
  0 & 0 \\
  -1 & 1 \\
 \end{bmatrix}
\end{align}
for standard polar codes ($Q=2^q=1$), un-optimized QC polar codes,
optimized QC polar codes, and irregular QC polar codes, respectively.
We denote a pruned polarization by a negative shift value.
As the first two cases do not satisfy the condition in (\ref{eq:girth4}) to eliminate girth-$4$,
the BER performance is worse than the other two cases.
By eliminating the cycle-$4$ loop, the QC polar codes can achieve a gain of $0.9$~dB without sacrificing any computational complexity.
We found that frozen bit locations are also important; specifically, no gain was achieved with $[u_1, u_3]$ being frozen.

\begin{figure}[t]
 \centering
 \includegraphics[width=\linewidth]{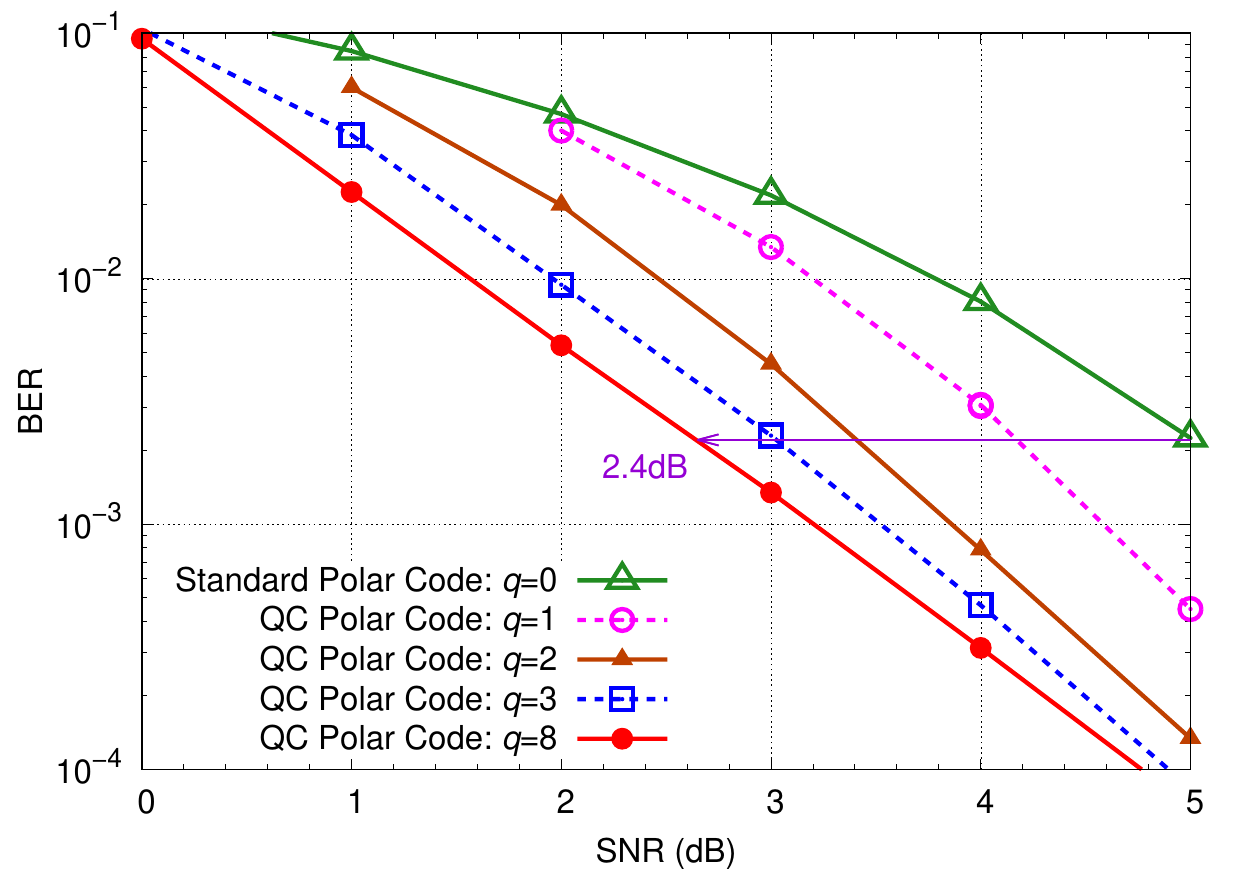}
 \caption{BER performance of $32$-iteration BP decoding for $4$-stage half-rate QC polar codes $(2^4,2^3,2^q)$.
 Frozen bit indications are $[1,1,1,1,1,1,1,0,1,0,0,0,0,0,0,0]$.}
 \label{fig:m4}
\end{figure}

The performance improvement can be more significant for a deeper stage $n$ and larger lifting size $Q$.
We plot the BER performance for half-rate $4$-stage QC polar codes $(2^4,2^3,2^q)$.
We can see that the increase of the lifting size $Q=2^q$ can significantly improve performance by up to $2.4$~dB gain over the standard polar codes.
It should be noted that the per-bit complexity is identical for all of these QC polar codes regardless of the lifting size $Q$;
specifically, $Q$ parallel decoding of short polar codes requires a total complexity on the order of $Q\times \mathcal{O}[n2^{n-1}]$ for a total codeword length of $Q\times 2^n$ bits.
For $Q=256$, our girth design method could remove all short cycles up to $6$.
The designed shift base matrix is written in (\ref{eq:shift16}), and also depicted in Fig.~\ref{fig:polar16}.

\begin{figure}[t]
 \centering
 \includegraphics[width=\linewidth]{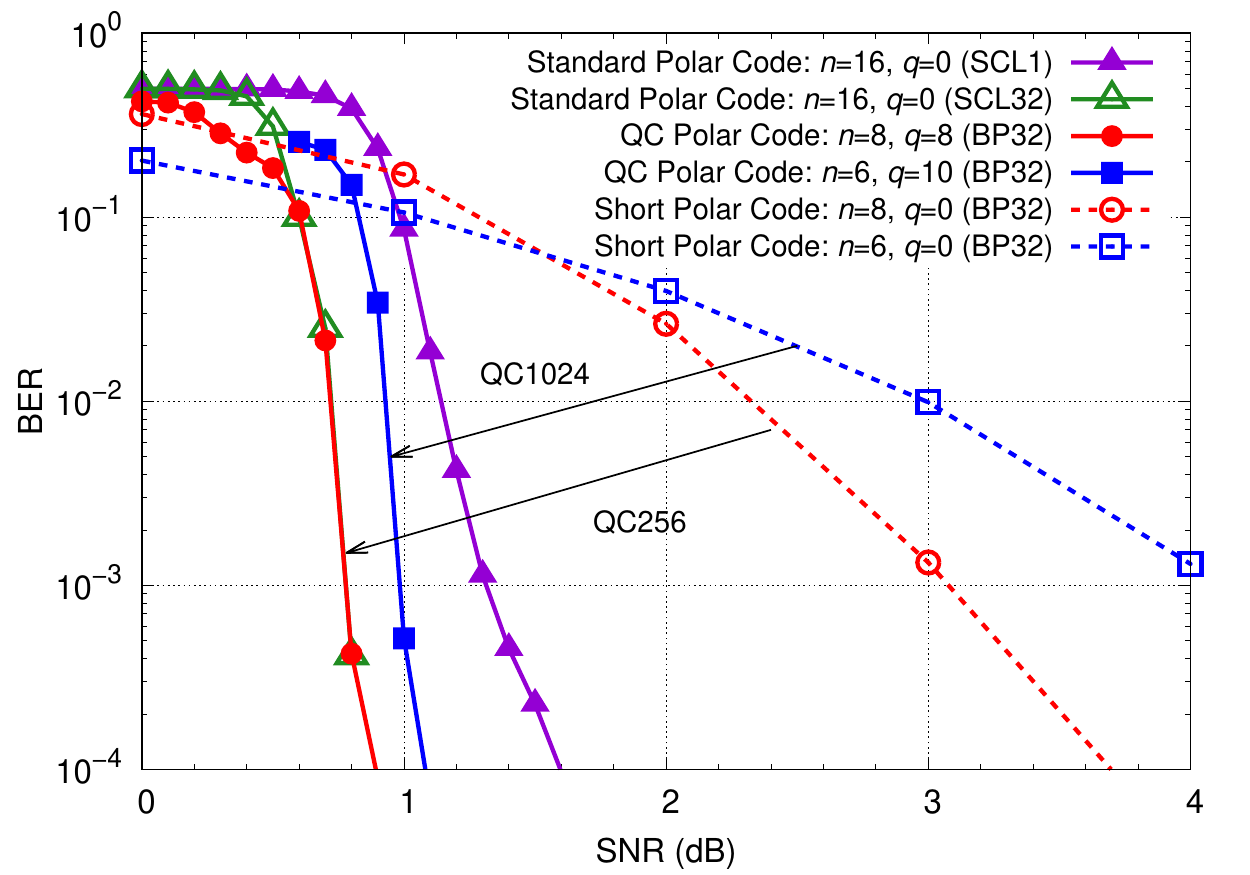}
 \caption{BER performance for long half-rate QC polar codes $(2^n, 2^{n-1}, 2^q)$ of block length $N=2^{n+q}=2^{16}$.}
 \label{fig:n16}
\end{figure}

We next demonstrate that our QC polar codes using shallow polarization stages can compete
against long standard polar codes with deeper polarization stages.
Fig.~\ref{fig:n16} shows the BER performance of $16$-stage standard polar codes and $6$-/$8$-stage QC polar codes for a total block length of $N=2^{16}$ bits.
We also present the shallow $6$-/$8$-stage polar codes without protograph lifting.
Noticeably, shallow $6$-stage QC polar codes with $Q=1024$ parallel BP decoding can outperform SC decoding of $16$-stage polar codes.
Furthermore, our $8$-stage QC polar codes with $Q=256$ can achieve performance competitive with state-of-the-art SCL decoding (with a list size of $L=32$) for the long $16$-stage polar codes.
In addition, it was verified that the QC polar codes can resolve the issues of BP decoding to offer comparable performance to SCL decoding with large list sizes.
These results are practically impactful because the encoding, decoding, and code design of shallower polar codes are much simpler and more efficient.

\subsection{Irregular Pruning}

\begin{figure}[t]
 \centering
 \includegraphics[width=\linewidth]{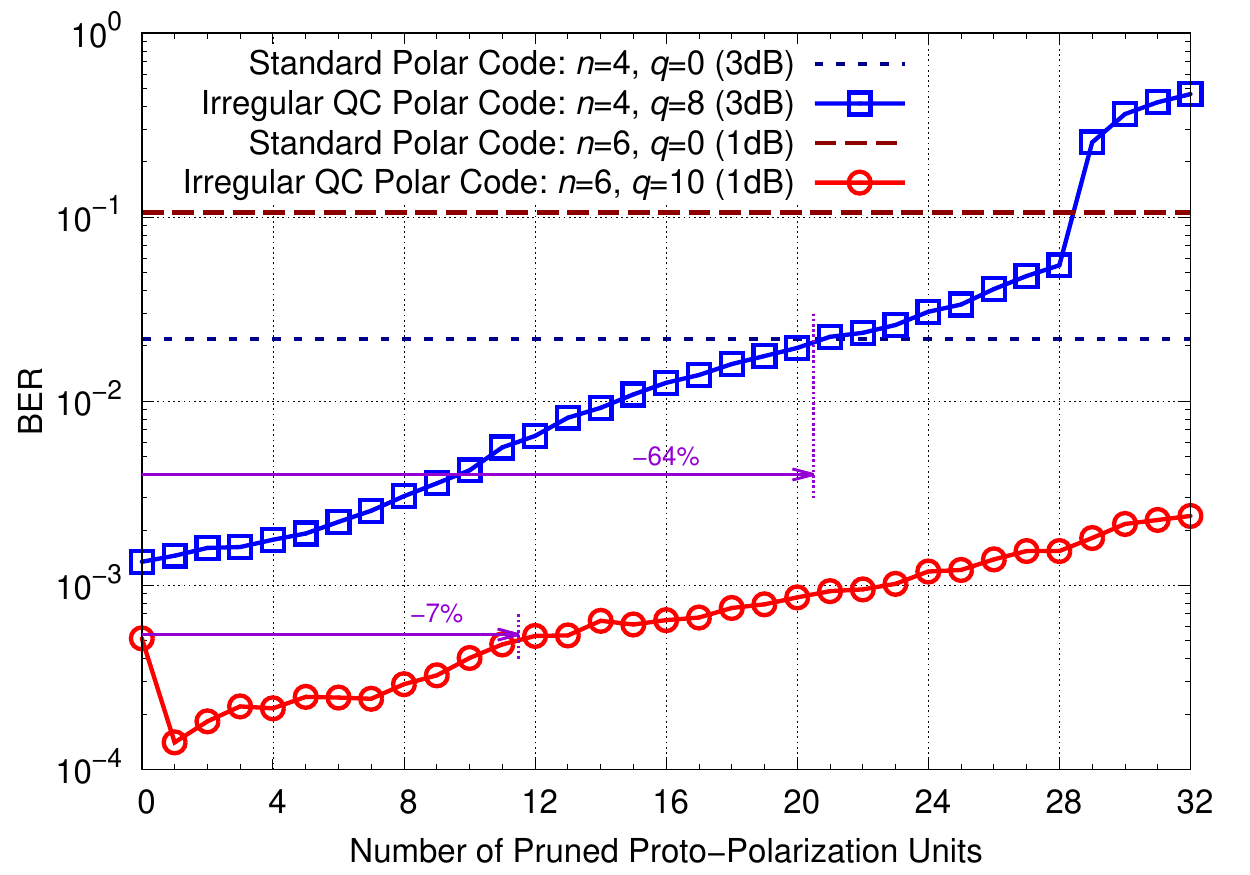}
 \caption{Impact of pruning proto-polarizations for irregular QC polar codes $(2^n,2^{n-1},2^q)$.}
 \label{fig:irreg}
\end{figure}

We finally investigate the irregular QC polar codes which deactivate polarization units.
As discussed, pruning proto-polarization units may also assist removing short cycles.
It was shown in\cite{Koike-JLT18} that the conventional irregular polar codes are often capable of reducing the encoding/decoding complexity, decoding latency, and even BER (due to improved Hamming weight distributions).
Fig.~\ref{fig:irreg} shows performance of QC polar codes $(2^n, 2^{n-1}, 2^q)$ when proto-polarization units are gradually pruned by a greedy algorithm\cite{Koike-JLT18}.
For $4$-stage polar codes, the performance degrades as the number of inactive polarizations increases.
Nevertheless, the QC polar codes with $Q=256$ are still better than the standard polar codes ($Q=1$) until $64$\% of the polar units are removed.
For $6$-stage polar codes with $Q=1024$ lifting, it was observed that pruning up to $7$\% of the proto-polarization units improves the BER performance over the regular counterpart.
In consequence, the irregular QC polar codes can further reduce the decoding complexity with potential performance improvement.

\subsection{Discussion}

Some major advantages of the proposed protograph polar codes are listed below:
\begin{itemize}
 \item The girth of polar codes can be increased significantly.
 \item The BP decoding can compete with SCL decoding.
 \item Multiple short polar encoders and decoders are implemented in a fully parallel fashion with no additional complexity besides circulant message exchanges.
 \item It realizes a low computational complexity equivalent to $Q$-fold shorter polar codes.
 \item Shallow polarization offers comparable performance to deeper polarization.
 \item Code design is simpler using shallower polarization.
 \item There is a higher flexibility in codeword lengths of non powers-of-two, i.e., $N=2^n Q$.
 \item Irregular polarization is straightforward to apply with the shift value matrix design.
 \item Well-established techniques such as girth design and P-EXIT from LDPC codes are applicable.
\end{itemize}

We however found that the recent BP list decoding\cite{Elkelesh-BPL} was not compatible with the QC polar codes as it is.
As we focused on the proof-of-principle study in this paper, there remain many research directions, including extensions to BP list decoding, systematic encoding, nonbinary codes, systematic circulant shift design, BP scheduling optimization, and multi-weight permutation.
In particular, it is interesting to consider inhomogeneous polar codes, e.g., non-identical frozen bit locations across $Q$ polar codes, and circulant permutations among different polarization stages.
We also note that our QC polar codes are similar to polar product codes\cite{Koike-ICC18} in the sense that parallel short polar codes are coupled, but the fundamental difference lies in its mechanism of coupling (QC polar codes do not need additional row/column polar codes but computation-free circulant).
Also, mixture-kernel and nonbinary polar codes are analogous to the QC polar codes in the sense that a single polarization unit processes multiple bits at once in parallel.
However, nonbinary polar codes require additional complexity and there is a limited flexibility to choose a Galois field size $Q$.
More importantly, $Q$-ary polar codes have only $\log_2(Q)$ bits parallelism, whereas fully $Q$-parallel encoding is possible in the QC polar codes.
We believe that the protograph design for generalized LDGM (including QC polar codes) will stimulate the research community.

\section{Conclusions}

We proposed a novel class of polar codes called QC polar codes,
which replicate parallel short polar encoders and decoders with circulant permutations to exchange intermediate messages among them.
We developed a protograph-based design method to optimize the girth to achieve high coding gain.
By removing short cycles, the proposed QC polar codes can outperform the standard polar codes.
We believe that the QC polar coding offers a breakthrough to resolve the long-standing issue that BP decoding performs poorly for conventional polar codes due to the inherent girth-4.
It was demonstrated that the QC polar codes with shallow polarization can achieve the state-of-the-art performance of deeper polar codes at a considerably reduced complexity.
The QC polar codes are hardware friendly as highly parallel encoding/decoding is feasible with a reduced polarization stage.
We also evaluated the impact of irregular QC polar codes, which can further decrease the complexity and BER for some cases.
For the proposed protograph codes, we addressed a number of interesting extensions for future research directions.

\section*{Acknowledgment}

We would like to thank
Drs. David S. Millar, Keisuke Kojima, and Kieran Parsons at MERL for their useful discussion.

\end{document}